\documentclass[twoside,twocolumn,english,prl,a4paper,showpacs,superscriptaddress]{revtex4}
\usepackage[T1]{fontenc}
\usepackage[latin9]{inputenc}
\usepackage{babel}
\usepackage{color}
\usepackage{ulem}
\pdfoutput=1
\usepackage{amsmath}
\usepackage{graphicx}
\usepackage{amssymb}
\usepackage{bbm}
\usepackage[unicode=true, pdfusetitle,
 bookmarks=true,bookmarksnumbered=false,bookmarksopen=false,
 breaklinks=false,pdfborder={0 0 1},backref=false,colorlinks=true]
 {hyperref}

\makeatletter
\@ifundefined{textcolor}{}
{%
 \definecolor{BLACK}{gray}{0}
 \definecolor{WHITE}{gray}{1}
 \definecolor{RED}{rgb}{1,0,0}
 \definecolor{GREEN}{rgb}{0,1,0}
 \definecolor{BLUE}{rgb}{0,0,1}
 \definecolor{CYAN}{cmyk}{1,0,0,0}
 \definecolor{MAGENTA}{cmyk}{0,1,0,0}
 \definecolor{YELLOW}{cmyk}{0,0,1,0}
 }

\newcommand{\otext}[1]{{\color{GREEN}}}

\usepackage{pdfsync}

\makeatother

\begin{document}

\title{QND Measurement of Large-Spin Ensembles by Dynamical Decoupling}
\author{M. Koschorreck}
    \email{marco.koschorreck@icfo.es}
    \affiliation{ICFO-Institut de Ciencies Fotoniques, 08860 Castelldefels (Barcelona), Spain}

\author{M. Napolitano}
    \affiliation{ICFO-Institut de Ciencies Fotoniques, 08860 Castelldefels (Barcelona), Spain}

\author{B. Dubost}
    \affiliation{ICFO-Institut de Ciencies Fotoniques, 08860 Castelldefels (Barcelona), Spain}
   \affiliation{Laboratoire Mat\'{e}riaux et Ph\'{e}nom\`{e}nes Quantiques, Universit\'{e} Paris Diderot et CNRS, \\UMR 7162, B\^{a}t. Condorcet, 75205 Paris Cedex 13, France}

\author{M. W. Mitchell}
  \affiliation{ICFO-Institut de Ciencies Fotoniques, 08860 Castelldefels (Barcelona), Spain}
  
\begin{abstract}
Quantum non-demolition (QND) measurement of collective variables by off-resonant
optical probing has the ability to create entanglement and squeezing
in atomic ensembles. Until now, this technique has been applied to
real or effective spin one-half systems. We show theoretically that
the build-up of Raman coherence prevents the naive application
of this technique to larger spin atoms, but that dynamical decoupling can be used 
to recover the ideal QND behavior.  We experimentally demonstrate
dynamical decoupling by using a two-polarization probing technique.
The decoupled QND measurement achieves a sensitivity  5.7(6)
dB better than the spin projection noise. 
\end{abstract}
\maketitle

Quantum non-demolition measurement plays a central role in quantum networking 
and quantum metrology for its ability to simultaneously detect and 
generate non-classical quantum states.  The original proposal by Braginsky \cite{Braginsky1974UFNv114p41} 
in the context of gravitational wave detection has been generalized to the optical \cite{Poizat1994APv19p265,Holland1990PRAv42p2995}, 
atomic \cite{Kuzmich1998ELv42p481} and nano-mechanical \cite{Ruskov2005PRBv71p235407} domains.  
In the atomic domain, QND by dispersive optical probing of spins or pseudo-spins
has been demonstrated using ensembles of cold atoms on a clock transition
\cite{Windpassinger2009MSTv20p55301, Schleier-Smith2010PRLv104p73604}, and with polarization variables
\cite{Takano2009PRLv102p33601,Koschorreck2010PRLv104p93602}, but thus 
far only with real or effective spin-1/2 systems. 


QND measurement of larger spin systems offers a metrological advantage, e.g., 
 in magnetometry \cite{Geremia2005PRLv94p203002}, and may be essential
 for the detection of different quantum phases of degenerate
atomic gases that intrinsically rely on large-spin systems \cite{Eckert2007NPv4p50,Eckert2007PRLv98p100404,Roscilde2009NJPv11p55041}.
Dispersive interactions with large-spin atoms are complicated by the presence of non-QND-type
terms in the effective Hamiltonian describing the interaction \cite{Geremia2006PRAv73p42112,Madsen2004PRAv70p52324, Echaniz2005JOBv7p548}.  
As we show, and contrary to what has often been assumed \cite{Kuzmich2000PRLv85p1594,Eckert2007NPv4p50,Eckert2007PRLv98p100404,Roscilde2009NJPv11p55041}, 
these terms spoil the QND performance, even in the large-detuning limit.   The non-QND terms introduce noise 
into the measured variable, or equivalently decoherence into the atomic state.  
The problem is serious for both large and small ensembles, so that naive application of dispersive 
probing fails for several of the above-cited proposals.  

We approach this problem using the methods of dynamical decoupling \cite{Viola1998PRAv58p2733, Viola1999PRLv82p2417,Facchi2005PRAv71p22302}, 
which allow us to effectively cancel the non-QND terms in the Hamiltonian while retaining the QND term. To our knowledge, this is the first application of this method to quantum non-demolition measurements.  Dynamical decoupling has been extensively applied in magnetic resonance \cite{Morton2008Nv455p1085, Biercuk2009Nv458p996}, used to suppress collisional decoherence in a thermal vapor \cite{Search2000PRLv85p2272}, to extend coherence times in solids \cite{ Taylor2008NPv4p810}, in Rydberg atoms \cite{Minns2006PRLv97p}, and with photon polarization \cite{Damodarakurup2009PRLv103p40502}. Other approaches include application of a static perturbation \cite{Smith2004PRLv93p163602,Fraval2004PRLv92p}. 

\global\long\def\rub{^{87}\mathrm{Rb}}

\global\long\def\gs{5S_{1/2}}

\global\long\def\es{5P_{3/2}}

\global\long\def\inlket#1#2{\left|\right.\!#1,\,#2\!\left.\right\rangle }

\global\long\def\ket#1#2{\left|#1,\,#2\right\rangle }

\global\long\def\sket#1{\left|#1\right\rangle }

\global\long\def\var#1{\mathrm{var}(#1)}
\global\long\def\cov#1{\mathrm{cov}(#1)}

\global\long\def\rubi{^{87}\mathrm{Rb}}

\global\long\def\Gcremat{G_{\gamma-V}}

\global\long\def\Sx{\hat{S}_{\mathrm{\mathbf{x}}}}
\global\long\def\Sy{\hat{S}_{\mathrm{\mathbf{y}}}}
\global\long\def\Sz{\hat{S}_{\mathrm{\mathbf{z}}}}
\global\long\def\Jx{\hat{J}_{\mathbf{x}}}
\global\long\def\Jy{\hat{J}_{\mathrm{\mathbf{y}}}}
\global\long\def\Jz{\hat{J}_{\mathrm{\mathbf{z}}}}

\global\long\def\sx{\hat{s}_{\mathrm{\mathbf{x}}}}
\global\long\def\sy#1{\hat{s}_{\mathrm{\mathbf{y}}#1}}
\global\long\def\sz{\hat{s}_{\mathrm{\mathbf{z}}}}
\global\long\def\jx{\hat{j}_{\mathbf{x}}}
\global\long\def\jy{\hat{j}_{\mathrm{\mathbf{y}}}}
\global\long\def\jz{\hat{j}_{\mathrm{\mathbf{z}}}}
\global\long\def\Tx{\hat{T}_{\mathrm{\mathbf{x}}}}
\global\long\def\Ty{\hat{T}_{\mathrm{\mathbf{y}}}}
\global\long\def\Tz{\hat{T}_{\mathrm{\mathbf{z}}}}
\global\long\def\Kx{\hat{K}_{\mathrm{\mathbf{x}}}}
\global\long\def\Ky{\hat{K}_{\mathrm{\mathbf{y}}}}
\global\long\def\Kz{\hat{K}_{\mathrm{\mathbf{z}}}}

\global\long\def\mSx{S_{\mathrm{\mathbf{x}}}}
\global\long\def\mSy{S_{\mathrm{\mathbf{y}}}}
\global\long\def\mSz{S_{\mathrm{\mathbf{z}}}}
\global\long\def\mJx{J_{\mathbf{x}}}
\global\long\def\mJy{J_{\mathrm{\mathbf{y}}}}
\global\long\def\mJz{J_{\mathrm{\mathbf{z}}}}

\global\long\def\msx{s_{\mathrm{\mathbf{x}}}}
\global\long\def\msy{s_{\mathrm{\mathbf{y}}}}
\global\long\def\msz{s_{\mathrm{\mathbf{z}}}}
\global\long\def\mjx{j_{\mathbf{x}}}
\global\long\def\mjy{j_{\mathrm{\mathbf{y}}}}
\global\long\def\mjz{j_{\mathrm{\mathbf{z}}}}
\global\long\def\dexpect#1{\left\langle \right.\!#1\!\left.\right\rangle }
\global\long\def\var#1{\mathrm{var}(#1)}
\global\long\def\sexpect#1#2{\left\langle #2\right|#1\left|#2\right\rangle }
 \global\long\def\Fx{\hat{F}_{x}}
\global\long\def\Fy{\hat{F}_{y}}
\global\long\def\Fz{\hat{F}_{z}}
\global\long\def\fx#1{\hat{f}_{x,#1}}
\global\long\def\fy#1{\hat{f}_{y,#1}}
\global\long\def\fz#1{\hat{f}_{z,#1}}

\newcommand{\vh}{\hat{v}}
\newcommand{\wh}{\hat{w}}
\newcommand{\Jh}{\hat{J}}
\newcommand{\Sh}{\hat{S}}
\newcommand{\Kh}{\hat{K}}
\newcommand{\Th}{\hat{T}}
\newcommand{\Uh}{\hat{U}}
\newcommand{\Ubb}{\hat{U}_{b}}

\newcommand{\bS}{{\bf S}}
\newcommand{\bJ}{{\bf J}}
\newcommand{\bj}{{\bf j}}
\newcommand{\boldf}{{\bf f}}
\newcommand{\balpha}{{\bf \alpha}}
\newcommand{\supone}{^{({1})} }
\newcommand{\suptwo}{^{({2})} }
\newcommand{\supin}{^{({\rm in})} }
\newcommand{\supmid}{^{({\rm mid})} }
\newcommand{\supout}{^{({\rm out})} }
\newcommand{\suponein}{^{(1,{\rm in})} }
\newcommand{\suponeout}{^{(1,{\rm out})} }
\newcommand{\suptwoin}{^{(2,{\rm in})} }
\newcommand{\suptwoout}{^{(2,{\rm out})} }
\newcommand{\SNR}{S_{\rm SNR}}
\newcommand{\xycomm}{\hat{j}_{[\mathbf{x,y}]}}
\newcommand{\XYcomm}{\hat{J}_{[\mathbf{x,y}]}}
\newcommand{\iden}{\mathbbm{1}}
\newcommand{\expect}[1]{\left<#1\right>}


We consider an ensemble of spin-$f$ atoms interacting with a pulse of near-resonant polarized light.   As described in references \cite{Geremia2006PRAv73p42112, Madsen2004PRAv70p52324, Echaniz2005JOBv7p548}, the light and atoms interact by the effective Hamiltonian $\hat{H}_{\rm eff}$
\begin{equation}
 \tau {\hat{H}}_{\mathrm{eff}} =  G_1 \Sz \Jz+G_2 ( \Sx \Jx +  \Sy \Jy)\,\,, \label{eq:H_full}
 \end{equation}
where $\tau$ is the duration of the pulse and $G_{1,2}$ are coupling constants that depend on the atomic absorption cross section, the beam geometry, the detuning from resonance $\Delta$, and the hyperfine structure of the atom \cite{Kubasik2009PRAv79p43815}.  The atomic variables $\hat{\bJ}$ (described below) are collective spin and alignment operators.  The light is described by the Stokes operators $\hat{\bS}$ defined as  $\hat{S}_i \equiv \frac{1}{2}(\hat{a}_+^\dagger,\hat{a}_-^\dagger) \sigma_i (\hat{a}_+,\hat{a}_-)^T$, where the $\sigma_i$ are the Pauli matrices and $\hat{a}_\pm$ are annihilation operators for the temporal mode of the pulse and circular plus/minus polarization. Bold subscripts, e.g., $\mathbf{x}$, are used to label non-spatial directions for atomic and light variables.  The $G_1$ term describes a QND interaction, while the $G_2$ describes a more complicated coupling.  In the dispersive, i.e. far-detuned, regime, $G_1$ and $G_2$ scale as $\Delta^{-1}$ and $\Delta^{-2}$, respectively. It has sometimes been assumed that the $G_2$ terms can be neglected for sufficiently large $\Delta$, leaving an approximate QND interaction.  As we show below, this scaling argument fails, and the $G_2$ terms remain important. We note an important symmetry: $\hat{H}_{\rm eff}$ commutes with $\Sz + \Jz$, and is thus invariant under simultaneous rotation of $\hat{\bJ}$ and $\hat{\bS}$ about the $z$ axis.  

The atomic collective variables are $\hat{J}_k \equiv \sum_{i}^{N_{A}} \hat{j}^{(i)}_k$ where the superscript indicates the $i$-th atom and $\jx \equiv (\hat{f}_x^2 - \hat{f}_y^2 )/2$, $\jy \equiv (\hat{f}_x \hat{f}_y + \hat{f}_y \hat{f}_x)/2$,  $\jz \equiv \hat{f}_z/2$  and   $\xycomm  \equiv -i [\jx,\jy] =  \hat{f}_z (\hat{f}^2 -  \hat{f}_z^2 -1/2 ) $.   These obey commutation relations $[\jz,\jx] = i \jy$, $ [\jy,\jz] = i \jx$, $ [\jx,\jy] =  i \xycomm$.   For $f=1/2$, $\jx,\jy$ and $\xycomm$ vanish identically while for $f=1$, $\xycomm=\jz$ so that $\jx,\jy,$ and $\jz$ describe a pseudo-spin $\hat{\bf j}$.   




In the QND scenario, an initial coherent polarization state with $\dexpect{\hat{{\bf S}}} = (N_L/2,0,0)$ is passed through the ensemble and experiences a rotation due to the $G_1$ term such that the component $\Sy$ (the `meter' variable) indicates the value of $\Jz$ (the `system' variable). We assume that $\Jx = N_A/2$.  For a weak pulse, i.e., for $\dexpect{\hat{{\bf S}}} $ sufficiently small, we have the $\tau$-linear input-output relations $\hat{A}\supout = \hat{A}\supin - i \tau [\hat{A}\supin,\hat{H}_{\rm eff}]$.  Of specific interest are 
\newcommand{\spacer}{}

\begin{eqnarray}
\Jz\supout &=& \Jz\supin \spacer+ G_2 \Sx \Jy\supin -{G_2 \Sy\supin \Jx} \,\,,\label{JzInOut} \\
\Jy\supout &=& \Jy\supin - G_1 \Sz\supin \Jx  - G_2 \Sx \XYcomm \supin \,\,, \label{JyInOut} \\
\Sy\supout &=& \Sy\supin + G_1 \Sx \Jz\supin  -{G_2 \Sz\supin \Jy} \,\,,\label{SyInOut}
\end{eqnarray}
which describe the change in the system variable, its conjugate, and the meter variable.  
In the case of $f=1/2$, the $G_2$ terms vanish identically and we have a pure QND measurement: information about $\Jz$ enters $\Sy$ and there is a  back-action on $\Jy$, but not on $\Jz$.  
The input noise $\var{\Sy\supin} = \mSx/2$ limits the performance of the measurement, and corresponds to a spin sensitivity of $\delta \Jz^2 = (2 G_1^2 \Sx)^{-1}$.  For comparison, the projection noise of an ${\mathbf{x}}$-polarized spin state is $\var{\Jz} = \Jx/2$, so that projection noise sensitivity is achieved for $\Sx = (G_1^2 \Jx)^{-1} \equiv \SNR$. 

This ideal QND regime does not occur naturally except for $f=1/2$. In the interesting regime  $\Sx \approx \SNR$, we find that ${G_2 \Sx \Jy} \approx   \Jy(G_{2}/G_{1}^2)/\Jx$ is independent of $\Delta$, and cannot be neglected based on detuning.  
To get an order of magnitude, we note that for large detuning, $G_1 \approx {\sigma_0 \Gamma}/{4 A \Delta}$, $G_2 \approx G_1 \Delta_{\rm HFS}/\Delta$  where $\sigma_0$ is the on-resonance scattering cross-section, $A$ is the effective area of the beam, and $\Gamma$ and  $\Delta_{\rm HFS}$ are the natural linewidth and hyperfine splitting, respectively, of the excited states.  In terms of the on-resonance optical depth $d_0 \equiv \sigma_0 N_A / A$, we find $G_2/G_1^2 \mJx \approx 8 \Delta_{\rm HFS}/d_0 \Gamma$.  In a typical experiment with rubidium on the $D_2$ line, $\Delta_{\rm HFS}/\Gamma \sim 30$ and 
$d_0 \sim 50$ \cite{Kubasik2009PRAv79p43815}, so the contribution of this term is important.

%
In contrast, the last term in Eq. (\ref{JyInOut}) and  (\ref{SyInOut}), respectively,  contribute variances $\left<G_2^2 \Sy^2 \Jx^2\right>$ and $\left<G_2^2 \Sz^2 \Jy^2\right>$  which scale as $\Delta^{-2}$.  We will henceforth drop these terms.  



%


The system variable $\Jz$ is coupled to a degree of freedom, $\Jy$, which is neither system nor meter in the QND measurement.  This coupling introduces noise into the system variable, and decoherence into the state of the ensemble.   To remove the decoherence associated with this coupling $G_2 \Sx \Jy$, we adopt the strategy of ``bang-bang'' dynamical decoupling \cite{Viola1998PRAv58p2733, Viola1999PRLv82p2417, Facchi2005PRAv71p22302}.  In this method, a unitary $\Ubb$ and its inverse $\Ubb^\dagger$ are alternately and periodically applied to the system $p$ times during the evolution, so that the total evolution is $[\Ubb^\dagger \Uh_H(t/2p) \Ubb \Uh_H(t/2p)]^p$ where $\Uh_H(t)$ describes unitary evolution under $\hat{H}$ for a time $t$.  With this evolution, those system variables that are unchanged by $\Ubb$ continue to evolve under $\hat{H}$, while others are rapidly switched from one value to another, preventing coherent evolution.  For large $p$, the system evolves under a modified  Hamiltonian $\hat{H}' = \hat{P} \hat{H}$, where $\hat{P}$ projects onto the 
commutant (i.e., the set of operators which commute with) of $\{ \Ubb,\Ubb^\dagger \}$ \cite{Facchi2005PRAv71p22302}.  


To eliminate $G_2 (\Sx \Jx + \Sy \Jy)$, while keeping $G_1 \Sz \Jz$ we choose a $\Ubb$ which commutes with $\Jz$, but not with $\Jx$ or $\Jy$, namely a $\pi$ rotation about $\Jz$, $\Ubb = \exp[i \pi \Jz]$.  This leaves $\Jz$ unchanged, but inverts $\Jx$ and $\Jy$.  
By the symmetry of $\hat{H}_{\rm eff}$, this is equivalent to inverting $\Sx$ and $\Sy$, which suggests a practical implementation: probe with pulses of alternating $\Sx$, and define a `meter' variable taking into account the inversion of $\Sy$.  

We consider sequential interaction of the ensemble with a pair of pulses, with $\Sx^{(1)} = -\Sx^{(2)}=N_L/4p$.  We define also the new `meter' variable $S_y^{(\rm diff)} \equiv \Sy^{(1)} -\Sy^{(2)}$.  We describe the atomic variables before, between, and after the two pulses with superscripts $(\rm in),(mid),(out)$, respectively.  We apply Equations (\ref{JzInOut}-\ref{SyInOut}) to find:
\begin{eqnarray}
\Jz\supmid &=& \Jz\supin \spacer+ G_2 \Sx\supone \Jy\supin \label{JzInOutFP} \\
\Jy\supmid &=& \Jy\supin - G_1 \Sz\suponein \Jx  - G_2 \Sx\supone \XYcomm\supin  \label{JzInOutFP} \\
\Sy\suponeout &=& \Sy\suponein + G_1 \Sx\supone \Jz\supin   \label{SyInOutFP}
\end{eqnarray}
and
\begin{eqnarray}
\Jz\supout &=&  \Jz\supin  \label{JzInOutTP} \\ 
\Sy^{(\rm diff,out)} &=& \Sy^{(\rm diff,in)}  + 2G_1 \Sx\supone \Jz\supin  \label{SyInOutTP}
\end{eqnarray}
plus terms in $G_1 G_2 \Sx \Sz \Jx $, $G_2^2 \Sx^2 \XYcomm$ and $G_1 G_2 \Sx^2 \Jy$ which become negligible in the limit of large $p$.  The ideal QND form is recovered by the dynamical decoupling.

The presence of the $G_2$ term can be detected by noise scaling properties.  While in the ideal QND of Equations (\ref{JzInOutTP}),(\ref{SyInOutTP}) the variance of the system variable is $\propto \Jx$ giving a variance for the meter variable linear in $\Jx$, for the imperfect QND of Equations (\ref{JzInOut}) to (\ref{SyInOut}) this is not the case:  from Equation (\ref{JzInOutFP}), we see that $\Jy$ acquires a back-action variance $\propto \Jx^2$, which then is fed into the system variable by the $G_2$ term.  This additional $\Jx^2$ noise is also reflected in the meter variable, and provides a measurable indication of $G_2$.

We use the two-polarization decoupling technique to perform QND measurement on 
 an ensemble of $\sim10^{6}$ laser cooled $^{87}$Rb atoms in the $F=1$ ground state.  
In the atomic ensemble system, described in detail in reference  \cite{Kubasik2009PRAv79p43815}, 
 $\mu$s pulses interact with an elongated atomic cloud and are detected by a shot-noise-limited 
 polarimeter.  The experiment achieves projection noise limited 
sensitivity, as calibrated against a thermal spin state \cite{Koschorreck2010PRLv104p93602}. 

\begin{figure}[t!]
\includegraphics[width=1\columnwidth]{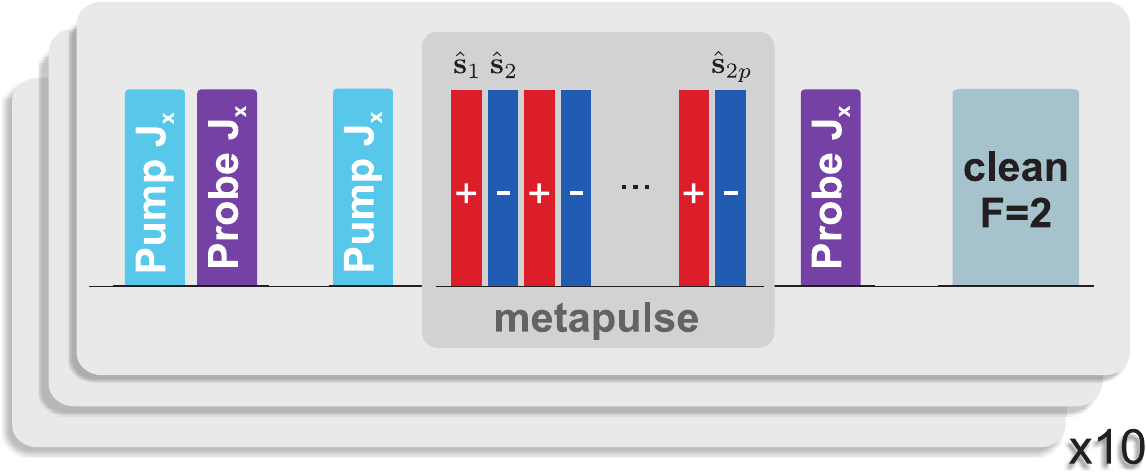}

\caption{\label{fig:TPP scheme}(color online) Experimental sequence for projection noise
measurement. The CSS is prepared once and its magnitude $\dexpect{ \Jx}$
is measured. This serves as a measure of the spin polarization prior
to the QND probing. We prepare the CSS a second time and assume it
has the same spin polarization as in the first preparation. The state
is probed with a train of pulses of alternating polarization.
Measuring the spin polarization after the QND measurement tells
us the amount of depolarization introduced in the QND probing. The
QND probing scatters a non-negligible fraction of atoms into $F=2$,
which are removed from the trap with resonant light in order to reduce
the number of atoms in the trap. The whole cycle is repeated 10 times
during one trap loading. }
\end{figure}

The experimental sequence is shown schematically in
Fig.~\ref{fig:TPP scheme}. In each measurement cycle the atom number $N_A$ is first measured by 
a dispersive atom-number measurement (DANM) \cite{Koschorreck2010PRLv104p93602}.  A $\Jx$-polarized
coherent spin state (CSS) is then prepared and probed with 
pulses of alternating polarization to find the QND signal $\Sy \equiv \sum_i \hat{s}_{\mathbf{y},i}\supout (-1)^{i+1}$.  
Immediately after, $\dexpect{\Jx}$ is measured to quantify depolarization of the sample and 
any atoms having made transitions to the $F=2$ manifold are 
removed from the trap, reducing $N_A$ for the next cycle and allowing a range of $N_A$ to be probed on a single loading. 
This sequence of state preparation and probing is repeated ten times for each loading of the trap.  
The trap is loaded 350 times to acquire statistics.  
 
The optical dipole trap, formed by a weakly-focused ($52\,\mu$m) beam of a Yb:YAG laser at $1030\,$nm
with $6\,$W of optical power, is loaded from a conventional
two stage magneto-optical trap (MOT) during $4\,$s. Sub-Doppler cooling
produces atom temperatures down to $25\,\mu$K as measured in the
dipole trap \cite{Kubasik2009PRAv79p43815}.  In the DANM, we prepare a $\Jx$-polarized CSS, i.e., all atoms in a 
 coherent superposition of hyperfine states $\left|\uparrow/\downarrow\right\rangle \equiv \left|F=1,m_F = \pm1\right\rangle $,
by optically pumping with vertically-polarized light tuned to the transition $F=1\rightarrow F'=1$,
while also applying repumping on the $F=2\rightarrow F'=2$ transition and a weak magnetic field along
$x$ to prevent spin precession. The atoms arrive to this dark state after scattering fewer than two photons on average.  
To measure $\dexpect{\Jx}$, we send ten circularly-polarized probe pulses, i.e., with $\dexpect{\Sz} = N_L/2$, tuned $190\,$MHz to the red of the transition
$F=1\rightarrow F'=0$.  Each pulse, of $1\,\mu$s duration, contains $2.6\times10^{6}$ photons and produces a 
signal $\dexpect{\Sy} \propto G_{2}\dexpect{ \Sz}\dexpect{ \Jx}$. 
The coherent state for the QND measurement is prepared in the same way, but in zero magnetic field.

To measure $ \Jz$, i.e., one half the population difference between
 $\left|\uparrow\right\rangle $ and $\left|\downarrow\right\rangle $, we send probe pulses of either 
vertical $\msx = n_L/2$ 
or horizontal $\msx = -n_L/2$ 
polarization through atomic sample and record their polarization rotation
as $\hat{s}_{\mathbf{y},i}^{\rm (out)}$. The number of individual probe pulses is $2p$ and
the total number of probe photons $N_{L}=2pn_{L}$.

\begin{figure}[t!]
\includegraphics[width=1\columnwidth]{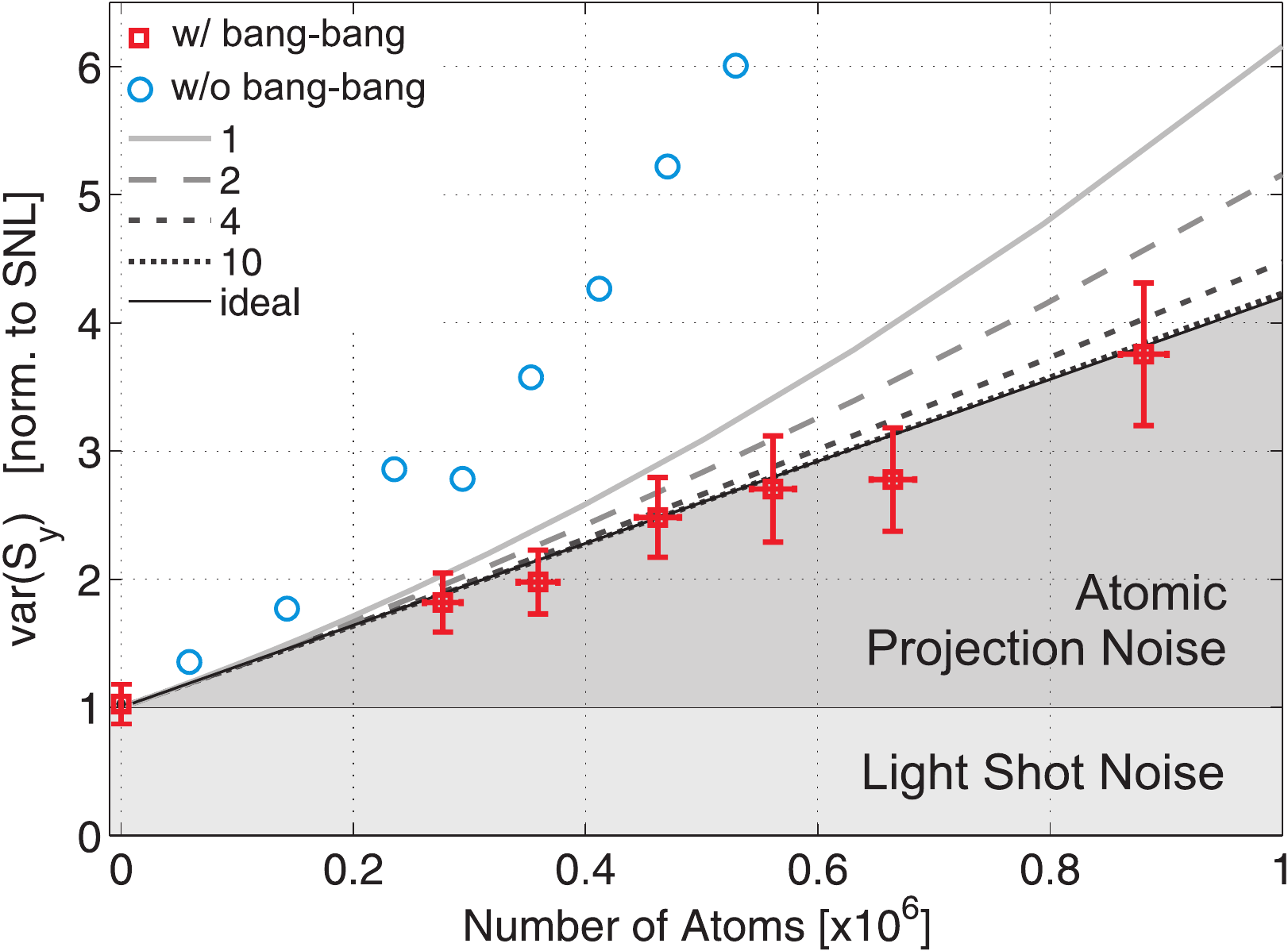}

\caption{\label{fig:Exp and Sim}(color online) 
Variance of polarimeter signal as a function of atom number, comparing naive probing, i.e., a single input polarization, to ``bang-bang'' dynamically-decoupled probing of different orders $p$.  Grey curves indicate simulation results for: naive probing (solid), and decoupled probing with $p=1$ (widely dashed), $p=2$ (dashed), and $p=5$ (dotted).
The black solid line shows the expected projection noise for
$p\rightarrow \infty$, or the ideal QND interaction $G2 = 0$.  All curves are calculated using the independently measured interaction strength $G_1 = 1.27(5)\times 10^{?7}$ and have no free parameters.  Red squares are measured data using dynamical decoupling with $p=5$.  Blue circles are measured data with naive probing.  Technical noise from laboratory fields dominates the naive probing results, and pushes them above the theoretical curve, while technical noise is suppressed in the dynamically-decoupled probing. }

\end{figure}


In Fig.~\ref{fig:Exp and Sim} we plot the measured noise versus atom number, which 
confirms the linear scaling characteristic of the QND measurement.  The black squares 
indicate the variance $\var{\Sy}$ normalized to the optical polarization noise, measured 
in the absence of atoms. Independent measurements confirm the polarimetry is 
shot-noise limited in this regime.  The black solid line is the expected projection noise 
scaling $4\var{ \Sy}/N_L=1+G_{1}^{2}N_{L}\var{ \Jz}$, calculated from the independently measured interaction strength $G_{1}$ and
number of probe photons $N_{L}=8\times10^{8}$. The QND measurement achieves 
projection-noise limited sensitivity, i.e., the measurement noise is  $5.7(6)\,$dB below
the projection noise.  

Also shown are results of covariance matrix calculations, following the techniques of reference 
\cite{Koschorreck2009JPBv42p9}, including loss and photon scattering.  The scenarios considered
include the naive QND measurement, i.e., with a single polarization, and the ``bang-bang'' or two-polarization
QND measurement, with $p=1,2,5$.  These show a rapid 
decrease in the quadratic component with increasing $p$.  
This confirms the removal of $G_2$ due to the dynamical decoupling.  Also included in these simulations 
is the term $ \Sy \Jy$ which introduces noise into $ \Jz$ proportional to $G_{2}^{2}\var{ \Sy}\dexpect{ \Jx}^{2}$. For our experimental
parameters this term leads to an increase of $\var{ \Jz}$ of less
then $2\,$\% and as noted above could be reduced with increased detuning.



The dynamical decoupling also suppresses technical
noise which would otherwise enter into $ \Jz$ through the interaction $G_{2}( \Sx \Jx+ \Sy \Jy)$.
An imperfect preparation of the atomic and and/or light state,
e.g., $\dexpect{ \Jy}\neq0$ or $\dexpect{ \Sy}\neq0$, would otherwise
be transferred into $ \Jz$. 

%

%

Using dynamical decoupling techniques, we have demonstrated optical quantum 
non-demolition measurement of a large-spin system.  We first identify an often-overlooked impediment to this goal:
the tensorial polarizability causes decoherence of the measured variable, and prevents (naive) QND measurement of small
ensembles.   We then identify an appropriate dynamical decoupling strategy to cancel the 
tensorial components of the effective Hamiltonian, and implement the strategy with an ensemble 
of $\sim 10^6$ cold $^{87}$Rb atoms and two-polarization probing.  The dynamically-decoupled QND 
measurement achieves a sensitivity $5.7(6)$ dB better than the projection noise level.  The technique will enable the use of 
large-spin ensembles in quantum metrology and quantum networking, and permit the QND measurement 
of exotic phases of large-spin condensed atomic gases.


We gratefully acknowledge fruitful discussions with Ivan H. Deutsch and Robert Sewell. This
work was funded by the Spanish Ministry of Science and Innovation
under the ILUMA project (Ref. FIS2008-01051) and the Consolider-Ingenio
2010 Project \textquoteleft{}\textquoteleft{}QOIT.\textquoteright{}\textquoteright{}

\bibliographystyle{apsrev_MK}
\bibliography{TPP}

\begin{thebibliography}{30}
\expandafter\ifx\csname natexlab\endcsname\relax\def\natexlab#1{#1}\fi
\expandafter\ifx\csname bibnamefont\endcsname\relax
  \def\bibnamefont#1{#1}\fi
\expandafter\ifx\csname bibfnamefont\endcsname\relax
  \def\bibfnamefont#1{#1}\fi
\expandafter\ifx\csname citenamefont\endcsname\relax
  \def\citenamefont#1{#1}\fi
\expandafter\ifx\csname url\endcsname\relax
  \def\url#1{\texttt{#1}}\fi
\expandafter\ifx\csname urlprefix\endcsname\relax\def\urlprefix{URL}\fi
\providecommand{\bibinfo}[2]{#2}
\providecommand{\eprint}[2][]{\url{#2}}

\bibitem[{\citenamefont{Braginsky and
  Vorontsov}(1974)}]{Braginsky1974UFNv114p41}
\bibinfo{author}{\bibfnamefont{V.~B.} \bibnamefont{Braginsky}}
  \bibnamefont{and} \bibinfo{author}{\bibfnamefont{Y.~I.}
  \bibnamefont{Vorontsov}}.
\newblock \bibinfo{journal}{Usp Fiz. Nauk} \textbf{\bibinfo{volume}{114}},
  \bibinfo{pages}{41} (\bibinfo{year}{1974}).

\bibitem[{\citenamefont{Poizat et~al.}(1994)\citenamefont{Poizat, Roch, and
  Grangier}}]{Poizat1994APv19p265}
\bibinfo{author}{\bibfnamefont{J.~P.} \bibnamefont{Poizat}},
  \bibinfo{author}{\bibfnamefont{J.~F.} \bibnamefont{Roch}}, \bibnamefont{and}
  \bibinfo{author}{\bibfnamefont{P.}~\bibnamefont{Grangier}}.
\newblock \bibinfo{journal}{Ann. Phys.-Paris} \textbf{\bibinfo{volume}{19}},
  \bibinfo{pages}{265} (\bibinfo{year}{1994}).

\bibitem[{\citenamefont{Holland et~al.}(1990)\citenamefont{Holland, Collett,
  Walls, and Levenson}}]{Holland1990PRAv42p2995}
\bibinfo{author}{\bibfnamefont{M.~J.} \bibnamefont{Holland}},
  \bibinfo{author}{\bibfnamefont{M.}~\bibnamefont{Collett}},
  \bibinfo{author}{\bibfnamefont{D.~F.} \bibnamefont{Walls}}, \bibnamefont{and}
  \bibinfo{author}{\bibfnamefont{M.~D.} \bibnamefont{Levenson}}.
\newblock \bibinfo{journal}{Phys. Rev. A} \textbf{\bibinfo{volume}{42}},
  \bibinfo{pages}{2995} (\bibinfo{year}{1990}).

\bibitem[{\citenamefont{Kuzmich et~al.}(1998)\citenamefont{Kuzmich, Bigelow,
  and Mandel}}]{Kuzmich1998ELv42p481}
\bibinfo{author}{\bibfnamefont{A.}~\bibnamefont{Kuzmich}},
  \bibinfo{author}{\bibfnamefont{N.~B.} \bibnamefont{Bigelow}},
  \bibnamefont{and} \bibinfo{author}{\bibfnamefont{L.}~\bibnamefont{Mandel}}.
\newblock \bibinfo{journal}{Europhys. Lett.} \textbf{\bibinfo{volume}{42}},
  \bibinfo{pages}{481} (\bibinfo{year}{1998}).

\bibitem[{\citenamefont{Ruskov et~al.}(2005)\citenamefont{Ruskov, Schwab, and
  Korotkov}}]{Ruskov2005PRBv71p235407}
\bibinfo{author}{\bibfnamefont{R.}~\bibnamefont{Ruskov}},
  \bibinfo{author}{\bibfnamefont{K.}~\bibnamefont{Schwab}}, \bibnamefont{and}
  \bibinfo{author}{\bibfnamefont{A.}~\bibnamefont{Korotkov}}.
\newblock \bibinfo{journal}{Phys. Rev. B} \textbf{\bibinfo{volume}{71}},
  \bibinfo{pages}{235407} (\bibinfo{year}{2005}).

\bibitem[{\citenamefont{Windpassinger et~al.}(2009)\citenamefont{Windpassinger,
  Kubasik, Koschorreck, Boisen, Kjaergaard, Polzik, and
  M{\"u}ller}}]{Windpassinger2009MSTv20p55301}
\bibinfo{author}{\bibfnamefont{P.~J.} \bibnamefont{Windpassinger}},
  \bibinfo{author}{\bibfnamefont{M.}~\bibnamefont{Kubasik}},
  \bibinfo{author}{\bibfnamefont{M.}~\bibnamefont{Koschorreck}},
  \bibinfo{author}{\bibfnamefont{A.}~\bibnamefont{Boisen}},
  \bibinfo{author}{\bibfnamefont{N.}~\bibnamefont{Kjaergaard}},
  \bibinfo{author}{\bibfnamefont{E.~S.} \bibnamefont{Polzik}},
  \bibnamefont{and} \bibinfo{author}{\bibfnamefont{J.~H.}
  \bibnamefont{M{\"u}ller}}.
\newblock \bibinfo{journal}{Meas. Sci. Technol.} \textbf{\bibinfo{volume}{20}},
  \bibinfo{pages}{055301} (\bibinfo{year}{2009}).

\bibitem[{\citenamefont{Schleier-Smith
  et~al.}(2010)\citenamefont{Schleier-Smith, Leroux, and
  Vuletic}}]{Schleier-Smith2010PRLv104p73604}
\bibinfo{author}{\bibfnamefont{M.~H.} \bibnamefont{Schleier-Smith}},
  \bibinfo{author}{\bibfnamefont{I.~D.} \bibnamefont{Leroux}},
  \bibnamefont{and} \bibinfo{author}{\bibfnamefont{V.}~\bibnamefont{Vuletic}}.
\newblock \bibinfo{journal}{Phys. Rev. Lett.} \textbf{\bibinfo{volume}{104}},
  \bibinfo{pages}{073604} (\bibinfo{year}{2010}).

\bibitem[{\citenamefont{Takano et~al.}(2009)\citenamefont{Takano, Fuyama,
  Namiki, and Takahashi}}]{Takano2009PRLv102p33601}
\bibinfo{author}{\bibfnamefont{T.}~\bibnamefont{Takano}},
  \bibinfo{author}{\bibfnamefont{M.}~\bibnamefont{Fuyama}},
  \bibinfo{author}{\bibfnamefont{R.}~\bibnamefont{Namiki}}, \bibnamefont{and}
  \bibinfo{author}{\bibfnamefont{Y.}~\bibnamefont{Takahashi}}.
\newblock \bibinfo{journal}{Phys. Rev. Lett.} \textbf{\bibinfo{volume}{102}},
  \bibinfo{pages}{033601} (\bibinfo{year}{2009}).

\bibitem[{\citenamefont{Koschorreck et~al.}(2010)\citenamefont{Koschorreck,
  Napolitano, Dubost, and Mitchell}}]{Koschorreck2010PRLv104p93602}
\bibinfo{author}{\bibfnamefont{M.}~\bibnamefont{Koschorreck}},
  \bibinfo{author}{\bibfnamefont{M.}~\bibnamefont{Napolitano}},
  \bibinfo{author}{\bibfnamefont{B.}~\bibnamefont{Dubost}}, \bibnamefont{and}
  \bibinfo{author}{\bibfnamefont{M.~W.} \bibnamefont{Mitchell}}.
\newblock \bibinfo{journal}{Phys. Rev. Lett.} \textbf{\bibinfo{volume}{104}},
  \bibinfo{pages}{093602} (\bibinfo{year}{2010}).

\bibitem[{\citenamefont{Geremia et~al.}(2005)\citenamefont{Geremia, Stockton,
  and Mabuchi}}]{Geremia2005PRLv94p203002}
\bibinfo{author}{\bibfnamefont{J.~M.} \bibnamefont{Geremia}},
  \bibinfo{author}{\bibfnamefont{J.~K.} \bibnamefont{Stockton}},
  \bibnamefont{and} \bibinfo{author}{\bibfnamefont{H.}~\bibnamefont{Mabuchi}}.
\newblock \bibinfo{journal}{Phys. Rev. Lett.} \textbf{\bibinfo{volume}{94}},
  \bibinfo{pages}{203002} (\bibinfo{year}{2005}).

\bibitem[{\citenamefont{Eckert et~al.}(2007{\natexlab{a}})\citenamefont{Eckert,
  Romero-Isart, Rodriguez, Lewenstein, Polzik, and
  Sanpera}}]{Eckert2007NPv4p50}
\bibinfo{author}{\bibfnamefont{K.}~\bibnamefont{Eckert}},
  \bibinfo{author}{\bibfnamefont{O.}~\bibnamefont{Romero-Isart}},
  \bibinfo{author}{\bibfnamefont{M.}~\bibnamefont{Rodriguez}},
  \bibinfo{author}{\bibfnamefont{M.}~\bibnamefont{Lewenstein}},
  \bibinfo{author}{\bibfnamefont{E.~S.} \bibnamefont{Polzik}},
  \bibnamefont{and} \bibinfo{author}{\bibfnamefont{A.}~\bibnamefont{Sanpera}}.
\newblock \bibinfo{journal}{Nature Physics} \textbf{\bibinfo{volume}{4}},
  \bibinfo{pages}{50} (\bibinfo{year}{2007}{\natexlab{a}}).

\bibitem[{\citenamefont{Eckert et~al.}(2007{\natexlab{b}})\citenamefont{Eckert,
  Zawitkowski, Sanpera, Lewenstein, and Polzik}}]{Eckert2007PRLv98p100404}
\bibinfo{author}{\bibfnamefont{K.}~\bibnamefont{Eckert}},
  \bibinfo{author}{\bibfnamefont{L.}~\bibnamefont{Zawitkowski}},
  \bibinfo{author}{\bibfnamefont{A.}~\bibnamefont{Sanpera}},
  \bibinfo{author}{\bibfnamefont{M.}~\bibnamefont{Lewenstein}},
  \bibnamefont{and} \bibinfo{author}{\bibfnamefont{E.~S.}
  \bibnamefont{Polzik}}.
\newblock \bibinfo{journal}{Phys. Rev. Lett.} \textbf{\bibinfo{volume}{98}},
  \bibinfo{pages}{100404} (\bibinfo{year}{2007}{\natexlab{b}}).

\bibitem[{\citenamefont{Roscilde et~al.}(2009)\citenamefont{Roscilde,
  Rodriguez, Eckert, Romero-Isart, Lewenstein, Polzik, and
  Sanpera}}]{Roscilde2009NJPv11p55041}
\bibinfo{author}{\bibfnamefont{T.}~\bibnamefont{Roscilde}},
  \bibinfo{author}{\bibfnamefont{M.}~\bibnamefont{Rodriguez}},
  \bibinfo{author}{\bibfnamefont{K.}~\bibnamefont{Eckert}},
  \bibinfo{author}{\bibfnamefont{O.}~\bibnamefont{Romero-Isart}},
  \bibinfo{author}{\bibfnamefont{M.}~\bibnamefont{Lewenstein}},
  \bibinfo{author}{\bibfnamefont{E.~S.} \bibnamefont{Polzik}},
  \bibnamefont{and} \bibinfo{author}{\bibfnamefont{A.}~\bibnamefont{Sanpera}}.
\newblock \bibinfo{journal}{New J. Phys.} \textbf{\bibinfo{volume}{11}},
  \bibinfo{pages}{055041} (\bibinfo{year}{2009}).

\bibitem[{\citenamefont{Geremia et~al.}(2006)\citenamefont{Geremia, Stockton,
  and Mabuchi}}]{Geremia2006PRAv73p42112}
\bibinfo{author}{\bibfnamefont{J.~M.} \bibnamefont{Geremia}},
  \bibinfo{author}{\bibfnamefont{J.~K.} \bibnamefont{Stockton}},
  \bibnamefont{and} \bibinfo{author}{\bibfnamefont{H.}~\bibnamefont{Mabuchi}}.
\newblock \bibinfo{journal}{Phys. Rev. A} \textbf{\bibinfo{volume}{73}},
  \bibinfo{pages}{42112} (\bibinfo{year}{2006}).

\bibitem[{\citenamefont{Madsen and M{\o}lmer}(2004)}]{Madsen2004PRAv70p52324}
\bibinfo{author}{\bibfnamefont{L.~B.} \bibnamefont{Madsen}} \bibnamefont{and}
  \bibinfo{author}{\bibfnamefont{K.}~\bibnamefont{M{\o}lmer}}.
\newblock \bibinfo{journal}{Phys. Rev. A} \textbf{\bibinfo{volume}{70}},
  \bibinfo{pages}{052324} (\bibinfo{year}{2004}).

\bibitem[{\citenamefont{de~Echaniz et~al.}(2005)\citenamefont{de~Echaniz,
  Mitchell, Kubasik, Koschorreck, Crepaz, Eschner, and
  Polzik}}]{Echaniz2005JOBv7p548}
\bibinfo{author}{\bibfnamefont{S.~R.} \bibnamefont{de~Echaniz}},
  \bibinfo{author}{\bibfnamefont{M.~W.} \bibnamefont{Mitchell}},
  \bibinfo{author}{\bibfnamefont{M.}~\bibnamefont{Kubasik}},
  \bibinfo{author}{\bibfnamefont{M.}~\bibnamefont{Koschorreck}},
  \bibinfo{author}{\bibfnamefont{H.}~\bibnamefont{Crepaz}},
  \bibinfo{author}{\bibfnamefont{J.}~\bibnamefont{Eschner}}, \bibnamefont{and}
  \bibinfo{author}{\bibfnamefont{E.~S.} \bibnamefont{Polzik}}.
\newblock \bibinfo{journal}{J. Opt. B} \textbf{\bibinfo{volume}{7}},
  \bibinfo{pages}{S548} (\bibinfo{year}{2005}).

\bibitem[{\citenamefont{Kuzmich et~al.}(2000)\citenamefont{Kuzmich, Mandel, and
  Bigelow}}]{Kuzmich2000PRLv85p1594}
\bibinfo{author}{\bibfnamefont{A.}~\bibnamefont{Kuzmich}},
  \bibinfo{author}{\bibfnamefont{L.}~\bibnamefont{Mandel}}, \bibnamefont{and}
  \bibinfo{author}{\bibfnamefont{N.~B.} \bibnamefont{Bigelow}}.
\newblock \bibinfo{journal}{Phys. Rev. Lett.} \textbf{\bibinfo{volume}{85}},
  \bibinfo{pages}{1594} (\bibinfo{year}{2000}).

\bibitem[{\citenamefont{Viola and Lloyd}(1998)}]{Viola1998PRAv58p2733}
\bibinfo{author}{\bibfnamefont{L.}~\bibnamefont{Viola}} \bibnamefont{and}
  \bibinfo{author}{\bibfnamefont{S.}~\bibnamefont{Lloyd}}.
\newblock \bibinfo{journal}{Phys. Rev. A} \textbf{\bibinfo{volume}{58}},
  \bibinfo{pages}{2733} (\bibinfo{year}{1998}).

\bibitem[{\citenamefont{Viola et~al.}(1999)\citenamefont{Viola, Knill, and
  Lloyd}}]{Viola1999PRLv82p2417}
\bibinfo{author}{\bibfnamefont{L.}~\bibnamefont{Viola}},
  \bibinfo{author}{\bibfnamefont{E.}~\bibnamefont{Knill}}, \bibnamefont{and}
  \bibinfo{author}{\bibfnamefont{S.}~\bibnamefont{Lloyd}}.
\newblock \bibinfo{journal}{Phys. Rev. Lett.} \textbf{\bibinfo{volume}{82}},
  \bibinfo{pages}{2417} (\bibinfo{year}{1999}).

\bibitem[{\citenamefont{Facchi et~al.}(2005)\citenamefont{Facchi, Tasaki,
  Pascazio, Nakazato, Tokuse, and Lidar}}]{Facchi2005PRAv71p22302}
\bibinfo{author}{\bibfnamefont{P.}~\bibnamefont{Facchi}},
  \bibinfo{author}{\bibfnamefont{S.}~\bibnamefont{Tasaki}},
  \bibinfo{author}{\bibfnamefont{S.}~\bibnamefont{Pascazio}},
  \bibinfo{author}{\bibfnamefont{H.}~\bibnamefont{Nakazato}},
  \bibinfo{author}{\bibfnamefont{A.}~\bibnamefont{Tokuse}}, \bibnamefont{and}
  \bibinfo{author}{\bibfnamefont{D.~A.} \bibnamefont{Lidar}}.
\newblock \bibinfo{journal}{Phys. Rev. A} \textbf{\bibinfo{volume}{71}},
  \bibinfo{pages}{22302} (\bibinfo{year}{2005}).

\bibitem[{\citenamefont{Morton et~al.}(2008)\citenamefont{Morton, Tyryshkin,
  Brown, Shankar, Lovett, Ardavan, Schenkel, Haller, Ager, and
  Lyon}}]{Morton2008Nv455p1085}
\bibinfo{author}{\bibfnamefont{J.~J.~L.} \bibnamefont{Morton}},
  \bibinfo{author}{\bibfnamefont{A.~M.} \bibnamefont{Tyryshkin}},
  \bibinfo{author}{\bibfnamefont{R.~M.} \bibnamefont{Brown}},
  \bibinfo{author}{\bibfnamefont{S.}~\bibnamefont{Shankar}},
  \bibinfo{author}{\bibfnamefont{B.~W.} \bibnamefont{Lovett}},
  \bibinfo{author}{\bibfnamefont{A.}~\bibnamefont{Ardavan}},
  \bibinfo{author}{\bibfnamefont{T.}~\bibnamefont{Schenkel}},
  \bibinfo{author}{\bibfnamefont{E.~E.} \bibnamefont{Haller}},
  \bibinfo{author}{\bibfnamefont{J.~W.} \bibnamefont{Ager}}, \bibnamefont{and}
  \bibinfo{author}{\bibfnamefont{S.~A.} \bibnamefont{Lyon}}.
\newblock \bibinfo{journal}{Nature} \textbf{\bibinfo{volume}{455}},
  \bibinfo{pages}{1085} (\bibinfo{year}{2008}).

\bibitem[{\citenamefont{Biercuk et~al.}(2009)\citenamefont{Biercuk, Uys,
  VanDevender, Shiga, Itano, and Bollinger}}]{Biercuk2009Nv458p996}
\bibinfo{author}{\bibfnamefont{M.~J.} \bibnamefont{Biercuk}},
  \bibinfo{author}{\bibfnamefont{H.}~\bibnamefont{Uys}},
  \bibinfo{author}{\bibfnamefont{A.~P.} \bibnamefont{VanDevender}},
  \bibinfo{author}{\bibfnamefont{N.}~\bibnamefont{Shiga}},
  \bibinfo{author}{\bibfnamefont{W.~M.} \bibnamefont{Itano}}, \bibnamefont{and}
  \bibinfo{author}{\bibfnamefont{J.~J.} \bibnamefont{Bollinger}}.
\newblock \bibinfo{journal}{Nature} \textbf{\bibinfo{volume}{458}},
  \bibinfo{pages}{996} (\bibinfo{year}{2009}).

\bibitem[{\citenamefont{Search and Berman}(2000)}]{Search2000PRLv85p2272}
\bibinfo{author}{\bibfnamefont{C.}~\bibnamefont{Search}} \bibnamefont{and}
  \bibinfo{author}{\bibfnamefont{P.}~\bibnamefont{Berman}}.
\newblock \bibinfo{journal}{Phys. Rev. Lett.} \textbf{\bibinfo{volume}{85}},
  \bibinfo{pages}{2272} (\bibinfo{year}{2000}).

\bibitem[{\citenamefont{Taylor et~al.}(2008)\citenamefont{Taylor, Cappellaro,
  Childress, Jiang, Budker, Hemmer, Yacoby, {\ldots}, and
  Lukin}}]{Taylor2008NPv4p810}
\bibinfo{author}{\bibfnamefont{J.}~\bibnamefont{Taylor}},
  \bibinfo{author}{\bibfnamefont{P.}~\bibnamefont{Cappellaro}},
  \bibinfo{author}{\bibfnamefont{L.}~\bibnamefont{Childress}},
  \bibinfo{author}{\bibfnamefont{L.}~\bibnamefont{Jiang}},
  \bibinfo{author}{\bibfnamefont{D.}~\bibnamefont{Budker}},
  \bibinfo{author}{\bibfnamefont{P.}~\bibnamefont{Hemmer}},
  \bibinfo{author}{\bibfnamefont{A.}~\bibnamefont{Yacoby}},
  \bibinfo{author}{\bibfnamefont{R.~L.~W.} \bibnamefont{{\ldots}}},
  \bibnamefont{and} \bibinfo{author}{\bibfnamefont{M.~D.} \bibnamefont{Lukin}}.
\newblock \bibinfo{journal}{Nature Physics} \textbf{\bibinfo{volume}{4}},
  \bibinfo{pages}{810} (\bibinfo{year}{2008}).

\bibitem[{\citenamefont{Minns et~al.}(2006)\citenamefont{Minns, Kutteruf,
  Zaidi, Ko, and Jones}}]{Minns2006PRLv97p}
\bibinfo{author}{\bibfnamefont{R.~S.} \bibnamefont{Minns}},
  \bibinfo{author}{\bibfnamefont{M.~R.} \bibnamefont{Kutteruf}},
  \bibinfo{author}{\bibfnamefont{H.}~\bibnamefont{Zaidi}},
  \bibinfo{author}{\bibfnamefont{L.}~\bibnamefont{Ko}}, \bibnamefont{and}
  \bibinfo{author}{\bibfnamefont{R.~R.} \bibnamefont{Jones}}.
\newblock \bibinfo{journal}{Phys. Rev. Lett.} \textbf{\bibinfo{volume}{97}}
  (\bibinfo{year}{2006}).

\bibitem[{\citenamefont{Damodarakurup et~al.}(2009)\citenamefont{Damodarakurup,
  Lucamarini, Giuseppe, Vitali, and Tombesi}}]{Damodarakurup2009PRLv103p40502}
\bibinfo{author}{\bibfnamefont{S.}~\bibnamefont{Damodarakurup}},
  \bibinfo{author}{\bibfnamefont{M.}~\bibnamefont{Lucamarini}},
  \bibinfo{author}{\bibfnamefont{G.~D.} \bibnamefont{Giuseppe}},
  \bibinfo{author}{\bibfnamefont{D.}~\bibnamefont{Vitali}}, \bibnamefont{and}
  \bibinfo{author}{\bibfnamefont{P.}~\bibnamefont{Tombesi}}.
\newblock \bibinfo{journal}{Phys. Rev. Lett.} \textbf{\bibinfo{volume}{103}},
  \bibinfo{pages}{40502} (\bibinfo{year}{2009}).

\bibitem[{\citenamefont{Smith et~al.}(2004)\citenamefont{Smith, Chaudhury,
  Silberfarb, Deutsch, and Jessen}}]{Smith2004PRLv93p163602}
\bibinfo{author}{\bibfnamefont{G.~A.} \bibnamefont{Smith}},
  \bibinfo{author}{\bibfnamefont{S.}~\bibnamefont{Chaudhury}},
  \bibinfo{author}{\bibfnamefont{A.}~\bibnamefont{Silberfarb}},
  \bibinfo{author}{\bibfnamefont{I.~H.} \bibnamefont{Deutsch}},
  \bibnamefont{and} \bibinfo{author}{\bibfnamefont{P.~S.}
  \bibnamefont{Jessen}}.
\newblock \bibinfo{journal}{Phys. Rev. Lett.} \textbf{\bibinfo{volume}{93}},
  \bibinfo{pages}{163602} (\bibinfo{year}{2004}).

\bibitem[{\citenamefont{Fraval et~al.}(2004)\citenamefont{Fraval, Sellars, and
  Longdell}}]{Fraval2004PRLv92p}
\bibinfo{author}{\bibfnamefont{E.}~\bibnamefont{Fraval}},
  \bibinfo{author}{\bibfnamefont{M.}~\bibnamefont{Sellars}}, \bibnamefont{and}
  \bibinfo{author}{\bibfnamefont{J.}~\bibnamefont{Longdell}}.
\newblock \bibinfo{journal}{Phys. Rev. Lett.} \textbf{\bibinfo{volume}{92}}
  (\bibinfo{year}{2004}).

\bibitem[{\citenamefont{Kubasik et~al.}(2009)\citenamefont{Kubasik,
  Koschorreck, Napolitano, de~Echaniz, Crepaz, Eschner, Polzik, and
  Mitchell}}]{Kubasik2009PRAv79p43815}
\bibinfo{author}{\bibfnamefont{M.}~\bibnamefont{Kubasik}},
  \bibinfo{author}{\bibfnamefont{M.}~\bibnamefont{Koschorreck}},
  \bibinfo{author}{\bibfnamefont{M.}~\bibnamefont{Napolitano}},
  \bibinfo{author}{\bibfnamefont{S.~R.} \bibnamefont{de~Echaniz}},
  \bibinfo{author}{\bibfnamefont{H.}~\bibnamefont{Crepaz}},
  \bibinfo{author}{\bibfnamefont{J.}~\bibnamefont{Eschner}},
  \bibinfo{author}{\bibfnamefont{E.~S.} \bibnamefont{Polzik}},
  \bibnamefont{and} \bibinfo{author}{\bibfnamefont{M.~W.}
  \bibnamefont{Mitchell}}.
\newblock \bibinfo{journal}{Phys. Rev. A} \textbf{\bibinfo{volume}{79}},
  \bibinfo{pages}{043815} (\bibinfo{year}{2009}).

\bibitem[{\citenamefont{Koschorreck and
  Mitchell}(2009)}]{Koschorreck2009JPBv42p9}
\bibinfo{author}{\bibfnamefont{M.}~\bibnamefont{Koschorreck}} \bibnamefont{and}
  \bibinfo{author}{\bibfnamefont{M.~W.} \bibnamefont{Mitchell}}.
\newblock \bibinfo{journal}{J. Phys. B} \textbf{\bibinfo{volume}{42}},
  \bibinfo{pages}{195502 (9pp)} (\bibinfo{year}{2009}).

\end{thebibliography}

\end{document}